\documentclass[aps,prl,twocolumn,showpacs]{revtex4}

\usepackage{mathptm}    
\usepackage{dcolumn}                    
\usepackage{bm}                         
\usepackage{graphicx}

\newcommand {\ba} {\begin{eqnarray}}
\newcommand {\ea} {\end{eqnarray}}

\usepackage{times}

\begin{document}

\title{Isotope effect and the role of phonon in the Fe-based superconductors}

\author{Yunkyu Bang}
\email[]{ykbang@chonnam.ac.kr} \affiliation{Department of Physics,
Chonnam National University, Kwangju 500-757, and Asia Pacific
Center for Theoretical Physics, Pohang 790-784, Korea}

\begin{abstract}
We studied the isotope effect of phonon in the Fe-based
superconductors using a phenomenological two band model for the
sign-changing s-wave ($\pm$s-wave) state. Within this mean-field
model, we showed that the large isotope effect is not inconsistent
with the $\pm$s-wave pairing state and its high transition
temperature. In principle, a large phonon isotope coefficient
$\alpha$  implies a large phonon coupling constant. However, the
asymmetric density of states (DOS) between two bands substantially
enhances the value of $\alpha$, so that a moderate value of the
phonon coupling constant ($\lambda_{\rm ph} \approx 0.4$) can
produce a very large value of $\alpha$ ($\approx 0.4$) as well as
a high transition temperature together with an antiferromagnet
(AFM)-induced interaction.
\end{abstract}

\pacs{74.20,74.20-z,74.50}

\date{\today}
\maketitle

Recent discovery of the Fe-based superconductors by Kamihara et
al. \cite{Kamihara,PhysToday08}, has greatly spurred the research
activity of unconventional superconductivity. Regarding the
pairing mechanism and symmetry of theses new superconducting (SC)
materials, there are already numerous experimental and theoretical
investigations. Since the first theoretical proposal of the
$\pm$s-wave state by Mazin et al., \cite{Mazin08} as a best
pairing state in the Fe-based superconductors (SCs), several
subsequent theoretical studies \cite{Kuroki,Eremin,DHLee,Bang08a}
supported this idea. Experiments such as ARPES \cite{Ding} and
penetration depth measurements \cite{pene} unanimously indicate a
full gap around the Fermi surfaces (FSs) consistent with a s-wave
gap state. Then the nuclear spin-lattice relaxation rate $1/T_1$
\cite{T1}, which is seemingly consistent with a nodal gap state
such as a d-wave gap, provided a strong evidence of the
sign-changing nature of the gaps on different bands, so that
actually strengthened the case of the $\pm$s-wave state
\cite{Bang08a,Bang08b,recent}.

As to the pairing glue, most researchers at the moment tend to
believe an electronic origin rather than a phonon origin paring
\cite{Kuroki,Bang08a,Eremin,DHLee,Tesanovic}. In particular, an
antiferromagnetic (AFM) correlation induced interaction appears to
be the most natural pairing glue in view of the common SDW
instability at around $\sim 150$ K and the overall phase diagram
with doping in this series of Fe pnictides \cite{neutron}. It is
shown by several authors that the AFM-induced potential, when
combined with the unique band structure (or the FS topology) of
the Fe pnictides, naturally leads to the $\pm$s-wave state as the
best pairing solution \cite{Kuroki,Bang08a,Eremin,DHLee}.

However, Liu et al. \cite{isotope} recently measured isotope
effect on $T_c$ with a substitution of $^{56}$Fe by $^{54}$Fe in
Ba$_{1-x}$K$_x$Fe$_2$As$_2$, and reported a unexpectedly large
isotope coefficient $\alpha$ ($\sim 0.4$). This observation, if
confirmed, is drastically perpendicular to the current line of
thought for the pairing mechanism of the Fe-based SCs. Theoretical
investigations about the electron-phonon coupling in these
materials are yet only a few. Boeri et al. \cite{Boeri} calculated
a very weak electron-phonon coupling constant ($\lambda_{\rm ph} <
0.2$) -- the value averaged over the Brillouin Zone (BZ). Eschrig
\cite{Eschrig} argued that the proper electron-phonon coupling is
not the averaged coupling constant but the one of a particular
phonon mode, i.e., in-plane Fe-breathing mode, which may have a
large electron-phonon coupling constant.

Besides theoretical investigations, the experimental fact is that
all Fe pnictides, either {\it R}FeAsO ({\it R}= La, Ce, and Nd) or
{\it A}KFe$_2$As$_2$ ({\it A}=Ba and Sr) compounds, display the
structural instability from tetragonal to orthorhombic symmetry
and it always occurs at temperatures very near the SDW transition
temperatures \cite{structure}. It implies, at least, two things:
(1) the lattice degrees of freedom (structural instability) and
the spin degrees of freedom (magnetic instability) are not
independent; (2) this structural instability is not a
Jahn-Teller-type instability of the Fe 3d-electrons but is closely
related with the metallicity of the Fe 3d-electrons
\cite{Tesanovic}. Therefore, the issue of the electron-phonon
coupling in the Fe pnictides needs to be further investigated.

With this motivation, in this paper, we studied the isotope effect
and the role of phonon using a phenomenological model for the
$\pm$s-wave pairing state. The details of the model can be found
in Ref.\cite{Bang08a} and here we briefly sketch the essential
ingredients for our purpose.
The model consists of two bands: one hole band centered around
$\Gamma$ point and one electron band centered around $M$ point in
the reduced Brillouin Zone (BZ) scheme, and has a phenomenological
pairing interaction, induced from an AFM correlation, hence
peaking around $(\pi, \pi)$ momentum exchange. In this paper, we
add a phonon interaction to this model. Because our purpose of
this paper is to study the isotope effect of phonon when the total
interaction gives rise to the $\pm$s-wave pairing state, we assume
only a general condition of the phonon interaction and vary the
basic parameters of the phonon interaction such as the coupling
strength $\lambda_{\rm ph}$ and the characteristic phonon
frequency $\omega_{\rm ph}$.

The Hamiltonian is written as

\begin{eqnarray}
H &=& \sum_{k \sigma} \epsilon_h (k) h^{\dag}_{k \sigma} h_{k
\sigma} + \sum_{k \sigma} \epsilon_e (k) e^{\dag}_{k \sigma} e_{k
\sigma} \nonumber \\
& & +\sum_{k k^{'} \uparrow \downarrow} V_{\rm AFM} (k,k^{'})
h^{\dag}_{k \uparrow} h^{\dag}_{-k \downarrow}
h_{k^{'} \downarrow}h_{-k^{'} \uparrow} \nonumber \\
& & +\sum_{k k^{'} \uparrow \downarrow} V_{\rm AFM} (k,k^{'})
e^{\dag}_{k \uparrow} e^{\dag}_{-k \downarrow} e_{k^{'}
\downarrow}e_{-k^{'}
\uparrow} \nonumber \\
& & + \sum_{k k^{'} \uparrow \downarrow} V_{\rm AFM} (k,k^{'})
h^{\dag}_{k \uparrow} h^{\dag}_{-k \downarrow} e_{k^{'}
\downarrow}e_{-k^{'}
\uparrow} \nonumber \\
& & +\sum_{k k^{'} \uparrow \downarrow} V_{\rm AFM} (k,k^{'})
e^{\dag}_{k \uparrow} e^{\dag}_{-k \downarrow} h_{k^{'}
\downarrow}h_{-k^{'} \uparrow} \nonumber \\
& & +\sum_{k k^{'} \uparrow \downarrow} V_{\rm ph} (k,k^{'})
h^{\dag}_{k \uparrow} h^{\dag}_{-k \downarrow}
h_{k^{'} \downarrow}h_{-k^{'} \uparrow} \nonumber \\
& & +\sum_{k k^{'} \uparrow \downarrow} V_{\rm ph} (k,k^{'})
e^{\dag}_{k \uparrow} e^{\dag}_{-k \downarrow} e_{k^{'}
\downarrow}e_{-k^{'} \uparrow}
\end{eqnarray}

\noindent where $\epsilon_{h,e} (k)$ are the dispersions of the
hole band and electron bands, respectively, representing two main
bands in the Fe pnictides. The details of the dispersions are not
important for our purpose except the density of states (DOS) of
each band, $N_h$ (hole band) and $N_e$ (electron band),
respectively. $h^{\dag}_{k \sigma}$ and $e^{\dag}_{k \sigma}$ are
the electron creation operators on the hole and the electron
bands, respectively. As mentioned previously, $V_{\rm AFM}
(k,k^{'})$ is the AFM-induced pairing potential, which is all
repulsive in momentum space, and $V_{\rm ph} (k,k^{'})$ is the
phonon interaction, which is all attractive in momentum space.

The minimum characteristics of the interactions to promote the
$\pm$s-wave gap solution are: $V_{\rm AFM} (k,k^{'})$, peaking
around $(\pi, \pi)$ momentum exchange, should have a stronger
interband interaction than the intraband one; and $V_{\rm ph}
(k,k^{'})$, being stronger for small momentum exchange, should
have a stronger intraband interaction than the interband one.
The latter requirement for the phonon interaction is already
included in the Hamiltonian by not including the interband terms
like $V_{\rm ph} (k,k^{'}) e^{\dag}_{k \uparrow} e^{\dag}_{-k
\downarrow} h_{k^{'} \downarrow}h_{-k^{'} \uparrow}$ and its
hermitian conjugate. This assumption of the phonon interaction is
indeed the property of the main phonons in Fe pnictides
\cite{Boeri}. If the phonon interaction were absolutely momentum
independent, it would have null effect for the $\pm$s-wave
pairing.

For simplicity of the analysis but without loss of generality, we
only need the FS-averaged interactions: for the AFM-induced
interactions such as $V_{\rm AFM} ^{he}= << V_{\rm AFM}(k_h,k_e)
>>_{k_h,k_e} $, $V_{\rm AFM} ^{hh}= << V_{\rm
AFM}(k_h,k'_h) >>_{k_h, k'_h}$, etc. and similarly for the phonon
interactions such as $V_{\rm ph} ^{hh}= - << V_{\rm ph}(k_h,k'_h)
>>_{k_h, k'_h}$ and $V_{\rm ph} ^{ee}= - <<
V_{\rm ph}(k_e,k'_e) >>_{k_e, k'_e}$. Notice that, in these
definitions, we absorbed the signs of the interactions and
therefore all $V_{\rm AFM} ^{ab}$ and $V_{\rm ph} ^{ab}$ are
positive values. Also assuming the $\pm$s-wave solution we fix the
signs of the s-wave gaps as $\Delta_h = |\Delta_h |$ on the hole
band and $\Delta_e  = - |\Delta_e |$ on the electron band,
respectively. The coupled $T_c$-equations are written as

\begin{eqnarray}
\Delta_h  &=&   -  \bigl[ V_{\rm AFM}^{hh}N_h \chi ^{\rm AFM} -
V_{\rm ph}^{hh} N_h \chi ^{\rm ph} \bigr] \Delta_h   \\ \nonumber
& & +   \bigl[ V_{\rm AFM}^{he} N_e \chi ^{\rm AFM} \bigr]
\Delta_e ,
\\ \nonumber
\Delta_e  &=&   -  \bigl[ V_{\rm AFM}^{ee} N_e \chi ^{\rm AFM}  -
V_{\rm ph}^{ee} N_e \chi ^{\rm ph} \bigr] \Delta_e \\ & & +
\bigl[ V_{\rm AFM}^{eh} N_h \chi ^{\rm AFM} \bigr] \Delta_h .
\end{eqnarray}

\noindent The pair susceptibilities $\chi ^{\rm AFM}$ and $\chi
^{\rm ph}$ at $T = T_c$ are defined as
\begin{eqnarray}
\chi (T_c) ^{\rm AFM, ph}&=& \int_{0} ^{\omega_{\rm AFM, ph}}
\frac{d \xi}{\xi} \tanh \Bigl[ \frac{\xi}{2 T_c} \Bigr]
\\ \nonumber & \approx &  \ln \Bigl[ \frac{1.14~
\omega_{\rm AFM, ph}}{T_c} \Bigr] .
\end{eqnarray}

\noindent where $\omega_{\rm AFM}$ and $\omega_{\rm ph}$ are the
cutoff frequencies of the AFM fluctuations and phonon,
respectively. The second expression is the well-known BCS
approximation valid only when $\omega_{\rm AFM, ph} \gg T_c$,
otherwise the first expression should be numerically calculated.
Equations (2) and (3) with Eq.(4) constitute the $T_c$-equation of
the $\pm$ s-wave state of the two band model, including phonon
pairing interaction as well as the AFM-induced pairing
interaction. Before we show the numerical results we can analyze a
simple case and gain general insight about the model.

In the case (we call it "symmetric case") that $N_h = N_e$, and
$V_{\rm AFM}^{ee} =V_{\rm AFM}^{hh}$ and $V_{\rm ph}^{ee} =V_{\rm
ph}^{hh}$ ($V_{\rm AFM}^{he} =V_{\rm AFM}^{eh}$ is always true),
we define the dimensionless pairing constants as follows,

\begin{eqnarray}
\lambda_{\rm AFM} ^{inter} &=& N_h V_{\rm AFM} ^{he}  = N_e V_{\rm AFM} ^{eh} \\
\lambda_{\rm AFM} ^{intra} &=& N_h V_{\rm AFM} ^{hh} =  N_e V_{\rm AFM} ^{ee}  \\
\lambda_{\rm ph}  &=& N_h V_{\rm ph} ^{hh} =  N_e V_{\rm ph}
^{ee}.
\end{eqnarray}

\noindent Then, the $T_c$-equations are simplified as

\begin{eqnarray}
\Delta_h (1 + \lambda_{\rm AFM} ^{intra} \chi ^{\rm AFM} -
\lambda_{\rm ph}  \chi ^{\rm ph}) &=&  \lambda_{\rm AFM} ^{inter}
\chi ^{\rm AFM}  \Delta_e ,
\\ \nonumber
\Delta_e (1 + \lambda_{\rm AFM} ^{intra} \chi ^{\rm AFM} -
\lambda_{\rm ph}  \chi ^{\rm ph})  &=&  \lambda_{\rm AFM} ^{inter}
\chi ^{\rm AFM}  \Delta_h  .
\end{eqnarray}

\noindent The above $T_c$-equations can be solved analytically
with the BCS approximation of the pair susceptibilities $\chi
^{\rm AFM}$ and $\chi ^{\rm ph}$, the second expression of Eq.(4),
as

\begin{equation}
T_c = 1.14 ~\omega_{\rm AFM} ^{\tilde{\lambda}_{\rm M}} \cdot
\omega_{\rm ph} ^{\tilde{\lambda}_{\rm ph}} \cdot
\exp^{-1/\lambda_{tot}}.
\end{equation}

\noindent where
\begin{eqnarray}
\lambda_{tot} &=& (\lambda_{\rm AFM} ^{inter} - \lambda_{\rm AFM}
^{intra}) + \lambda_{\rm ph} \\
\tilde{\lambda}_{\rm M} &=& (\lambda_{\rm AFM} ^{inter} -
\lambda_{\rm AFM} ^{intra}) / \lambda_{tot} \\
\tilde{\lambda}_{\rm ph} &=& \lambda_{\rm ph} / \lambda_{tot}.
\end{eqnarray}

\begin{figure}
\noindent
\includegraphics[width=100mm]{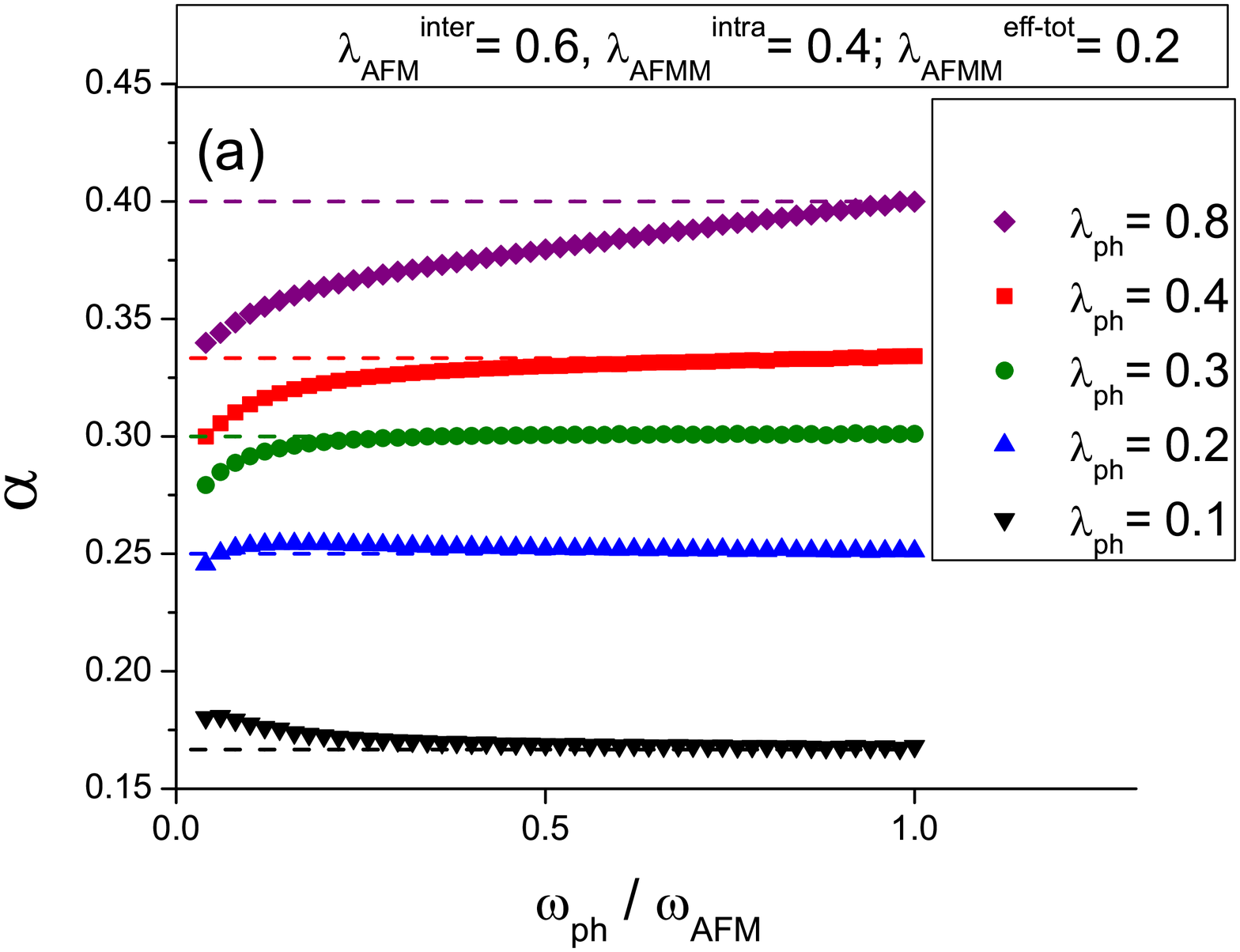}
\includegraphics[width=100mm]{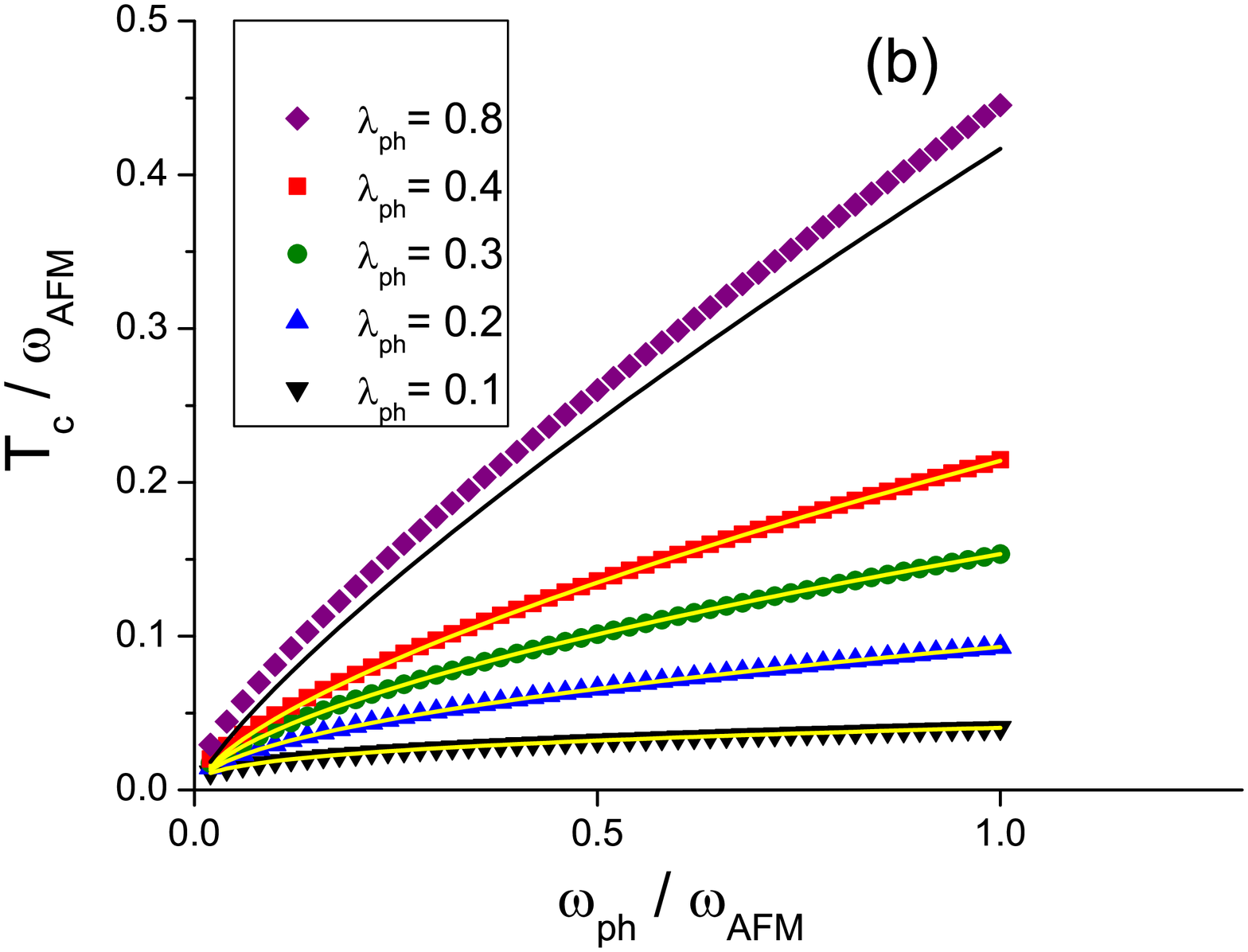}
\caption{(Color online)  Symmetric case ($N_h / N_e =1.0$). (a)
Isotope coefficients $\alpha$ as a function of the phonon cutoff
frequency $\omega_{\rm ph}/\omega_{\rm AFM}$ for different phonon
coupling constants $\lambda_{\rm ph} = 0.1, 0.2, 0.3, 0.4$ and
0.8. (b) Normalized $T_c$ vs $\omega_{\rm ph}/\omega_{\rm AFM}$
for the same parameters. Symbols are numerical calculations and
lines are the results of Eqs. (9) and (13). \label{fig1}}
\end{figure}

Equations (9) and (10) show that the AFM interaction and the
phonon interaction is additive  in the exponential form so that
$T_c$ can be dramatically boosted even with a small value of
$\lambda_{\rm ph}$ \cite{Bang-HTC}. We can also easily read the
phonon isotope coefficient $\alpha$  from Eq.(9) as
\begin{equation}
\alpha = \frac{1}{2}\frac{d \ln T_c}{d  \ln \omega_{\rm ph}} = 0.5
\times \frac{\lambda_{\rm ph}}{(\lambda_{\rm AFM} ^{inter} -
\lambda_{\rm AFM} ^{intra}) + \lambda_{\rm ph}}.
\end{equation}

\noindent This result conforms with a physical insight, i.e., a
large $\alpha$ value means a relatively large phonon coupling
compared to the total AFM pairing interaction $\lambda_{\rm AFM}
^{eff-tot}=(\lambda_{\rm AFM} ^{inter} - \lambda_{\rm AFM}
^{intra})$. However, Eq.(9) also shows that a large $\alpha$ value
does not necessarily mean the large phonon contribution to the
pairing energetics when $\omega_{\rm ph}$ is much smaller than
$\omega_{\rm AFM}$.

For the general cases where $N_h \neq N_e$, and $V_{\rm AFM}^{ee}
\neq V_{\rm AFM}^{hh}$ and $V_{\rm ph}^{ee} \neq V_{\rm ph}^{hh}$
("non-symmetric case"), it is not possible to find an analytic
solution of the $T_c$-equation. However, an inspection suggests us
to generalize Eq.(9) to the non-symmetric case. Equations (9) and
(13) can be used as good approximations for $T_c$ and $\alpha$
with the definitions of the effective dimensionless coupling
constants as follows,

\begin{eqnarray}
\lambda_{\rm AFM} ^{inter} &=& \sqrt{N_h N_e V_{\rm AFM} ^{he} V_{\rm AFM} ^{eh}} \\
\lambda_{\rm AFM} ^{intra} &=& \sqrt{N_h N_e V_{\rm AFM} ^{hh} V_{\rm AFM} ^{ee}}  \\
\lambda_{\rm ph}  &=& \sqrt{N_h N_e V_{\rm ph} ^{hh} V_{\rm ph}
^{ee}}.
\end{eqnarray}

\noindent In the following, we will show numerical results
directly obtained from Eqs. (2) and (3) and compare them with the
analytic formulas (9) and (13) both for the symmetric and
non-symmetric cases.

\begin{figure}
\noindent
\includegraphics[width=100mm]{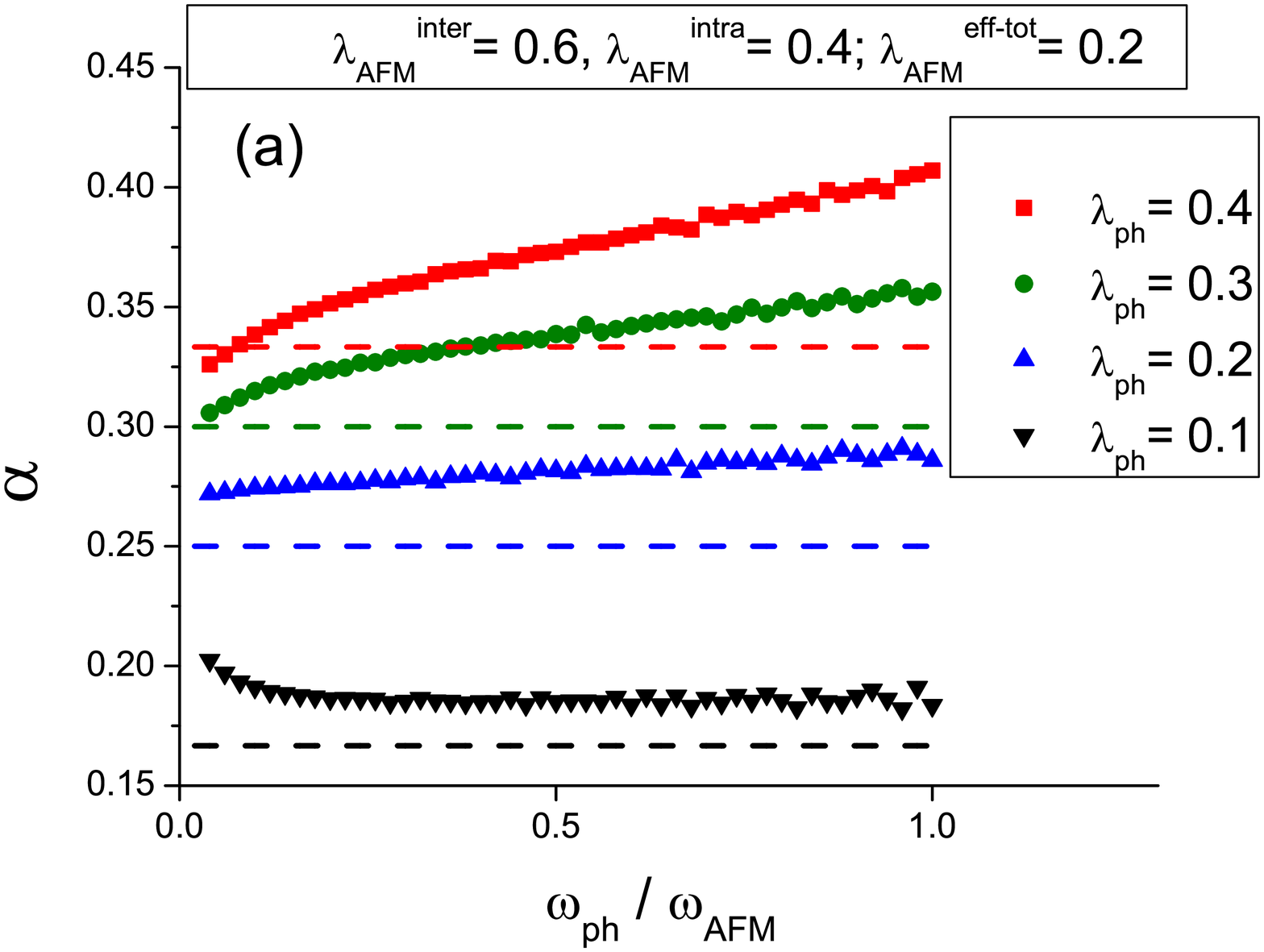}
\includegraphics[width=100mm]{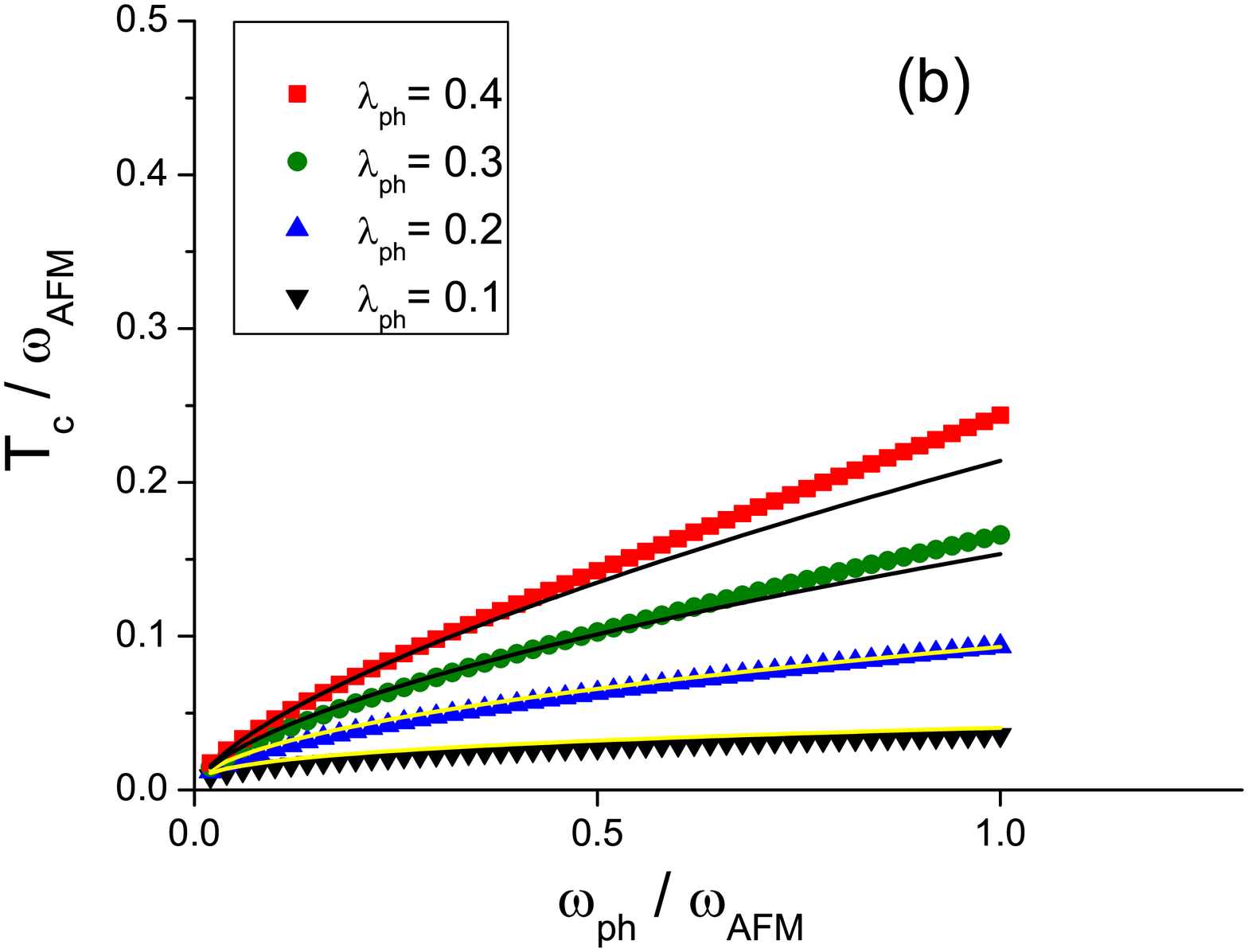}
\caption{(Color online)  Non-symmetric case ($N_h / N_e =3.0$).
(a) Isotope coefficients $\alpha$ as a function of the phonon
cutoff frequency $\omega_{\rm ph}/\omega_{\rm AFM}$ for different
phonon coupling constants $\lambda_{\rm ph} = 0.1, 0.2, 0.3$ and
0.4. (b) Normalized $T_c$ vs $\omega_{\rm ph}/\omega_{\rm AFM}$
for the same parameters. Symbols are numerical calculations and
lines are the results of Eqs. (9) and (13). \label{fig2}}
\end{figure}

Fig.1(a) shows the isotope coefficient $\alpha$ of the symmetric
case ($N_h = N_e $). Symbols are the numerical results of Eqs. (2)
and (3), and lines are the results of Eq.(13). All the energy
scales are normalized by the cutoff frequency of the
AFM-fluctuations $\omega_{\rm AFM}$ and we choose moderate
strength of dimensionless coupling constants of the AFM-induced
potential as $\lambda_{\rm AFM} ^{inter} = N_h V_{\rm AFM} ^{he} =
N_e V_{\rm AFM} ^{eh} =0.6$ and $\lambda_{\rm AFM} ^{intra} = N_h
V_{\rm AFM} ^{hh} = N_e V_{\rm AFM} ^{ee} =0.4$, which yields the
effective total magnetic interaction $\lambda_{\rm AFM}
^{eff-tot}=0.2$. Fig.1(a) shows that the numerically calculated
$\alpha$ (symbols) are in excellent agreement with the analytic
results of Eq.(13) when $\omega_{\rm ph} \gg T_c$. When this
condition is not fulfilled, deviations occurs at low $\omega_{\rm
ph}$ where the BCS approximation of the pair susceptibility
becomes poor.

The results of Fig.1(a) shows that a large phonon isotope
coefficient arises when the phonon coupling strength is much
stronger than the magnetic coupling strength and it is in accord
with a standard expectation. For example, in the symmetric case,
in order to obtain $\alpha \approx 0.4$ (the reported value by Liu
et al. \cite{isotope}), we need to have $\lambda_{\rm ph} = 4
\times \lambda_{\rm AFM} ^{eff-tot}$ which corresponds to the
result of $\lambda_{\rm ph}=0.8$ (purple diamond symbols). If this
is the case of reality, {\it then the superconductivity of the Fe
pnictides is a phonon driven SC and the AFM-fluctuations merely
act as a tipping agent to introduce the $\pi$ phase between two
bands widely separated in the BZ.} It is not an impossible
scenario, but we first need to find an evidence of such a strong
phonon coupling in Fe pnictides. If confirmed, the current
viewpoint about the pairing mechanism of the Fe-based SCs should
completely be changed.

In Fig.2 we show the results of the non-symmetric case ($N_h = 3
N_e$). We adjust the parameters to keep the dimensionless coupling
constants $\lambda_{\rm AFM} ^{inter} =0.6$ and $\lambda_{\rm AFM}
^{intra} =0.4$, the same values as in Fig.1. Here $\lambda_{\rm
AFM} ^{inter, intra}$ are calculated according to the generalized
formulas Eqs. (14) and (15). The phonon coupling constants
$\lambda_{\rm ph} =0.1, 0.2, 0.3, 0.4$ are also the values
according to Eq.(16). The behavior of $\alpha$ is very different
compared to the symmetric case: (1) there are systematic
deviations of the numerical (exact) results from the analytic
(approximate) formula Eq.(13) and the numerically calculated
values of $\alpha$ give substantially larger values than the ones
of the analytic formula. As a result, a moderate value of
$\lambda_{\rm ph} =0.4$ can cause a large isotope coefficient
$\alpha \approx 0.4$; (2) with increasing $\omega_{\rm ph}$, there
is no saturation of $\alpha$, in particular with large phonon
couplings. In contrast, the $T_c$ values are rather similar to the
symmetric case.

This seemingly inconsistent behavior between $T_c$ and $\alpha$
can be most easily understood by comparison of the $\lambda_{\rm
ph}=0.4$ results between the symmetric case (Fig.1) and the
non-symmetric case (Fig.2). $\alpha$ in the symmetric case quickly
saturates to its maximum value (red square symbols in Fig.1(a))
but $\alpha$ in the non-symmetric case gradually increases with
$\omega_{\rm ph}$ ( red square symbols in Fig.2(a)). As a result,
a large difference of the $\alpha$ value between the two cases is
possible when $\omega_{\rm ph}/ \omega_{\rm AFM} \rightarrow 1$
while maintaining only slightly enhanced $T_c$ (see red square
symbols in Fig.1(b) and in Fig.2(b)). Slightly enhanced $T_c$ is
due to the fact that the asymmetric phonon couplings on the hole
band and electron band are slightly more advantageous for pairing
than the equal phonon couplings on both bands when $\lambda_{\rm
ph}  = \sqrt{N_h N_e V_{\rm ph} ^{hh} V_{\rm ph} ^{ee}}$ is the
same.

In summary, using a minimal two band model with both the
AFM-induced interaction and the phonon interaction, which together
yields the $\pm$s-wave gap pairing state, we studied the phonon
isotope coefficient $\alpha$ for the symmetric case and the
non-symmetric case. We confirmed that a large value of $\alpha$
indicates a large value of the phonon coupling constant
$\lambda_{\rm ph}$ compared to the value of the AFM-induced
coupling constant $\lambda_{\rm AFM} ^{eff-tot}$. However, we
found that the large asymmetric ratio ($N_h / N_e$) of DOSs
between the hole and electron bands substantially enhances the
$\alpha$ value compared to the symmetric band case. As a result, a
relatively small value of the phonon coupling constant (say,
$\lambda_{\rm ph} =0.4$ when $\lambda_{\rm AFM} ^{eff-tot}$=0.2)
can yield a very large isotope coefficient $\alpha$=0.4.
Regardless of the symmetric or non-symmetric cases, $T_c$ is
strongly enhanced by the addition of two pairing interactions
$\lambda_{\rm ph}$ and $\lambda_{\rm AFM} ^{eff-tot}$ in the
exponential form. A similar result was obtained for the high-$T_c$
cuprates SCs \cite{Bang-HTC}. The possibly important role of
phonon(s) for the unconventional SCs discussed in this paper is a
very plausible scenario, in general. Although it is studied based
on a mean field theory, it is sufficient to demonstrate the
principle.
Further experimental study of the isotope effect in the Fe
pnictide SCs is a pressing task. If the large isotope effect is
confirmed, the importance of phonons for the unconventional SC
pairing in the correlated materials should be renewed.

{\it Acknowledgement -- } The author acknowledges useful
discussions  with S.-W. Cheong about the possible importance of
phonons in Fe pnictides. This work is supported by the KOSEF
through the Grant No. KRF-2007-521-C00081.

\end{document}